 \documentclass[iop,apjl, 12pt]{emulateapj}
\usepackage{xcolor, color, soul, natbib}
\usepackage{hyperref}
\hypersetup{  
    colorlinks = true, 
    citecolor = [cmyk]{1 .68 0 .12},
    urlcolor = [cmyk]{0 .95 1 0},
    linkcolor = [cmyk]{0 .95 1 0}
}
\usepackage[super]{nth} 
\usepackage{multirow, rotating,nth}

\newcommand{\myemail}{joe.llama@st-andrews.ac.uk}  

\newcommand{\rp}{$\textrm{R}_\textrm{p}/\textrm{R}_{\star} $\,}

\slugcomment{Accepted for publication in ApJ.}

\shorttitle{Transiting the Sun II: The impact of stellar activity on Lyman-$\alpha$ transits}
\shortauthors{Llama \& Shkolnik}
 
\begin{document}  
 
\title{Transiting the Sun II: The impact of stellar activity on Lyman-$\alpha$ transits}
\author{J. Llama$^{1,2}$ AND E.~L. Shkolnik$^{3, 1}$}
\affil{$^1$Lowell Observatory, 1400 W. Mars Hill Rd, Flagstaff, AZ 86001. USA \\
$^2$SUPA, School of Physics \& Astronomy. North Haugh. St Andrews. Fife. KY16 9SS. UK \\
$^3$ASU School of Earth and Space Exploration, Tempe, AZ 85287. USA} 
 
\email{Email: \myemail}

\begin{abstract}
High-energy observations of the Sun provide an opportunity to test the limits of our ability to accurately measure properties of transiting exoplanets in the presence of stellar activity. Here we insert transits of a hot Jupiter into continuous disk integrated data of the Sun in Lyman-$\alpha$ (Ly$\alpha$) from NASA's SDO/EVE instrument to assess the impact of stellar activity on the measured planet-to-star radius ratio (\rp). In 75\% of our simulated light curves we measure the correct radius ratio; however, incorrect values can be measured if there is significant short term variability in the light curve. The maximum measured value of \rp is $50\%$ larger than the input value, which is much smaller than the large Ly$\alpha$ transit depths that have been reported in the literature, suggesting that for stars with activity levels comparable to the Sun, stellar activity alone cannot account for these deep transits. We ran simulations without a transit and found that stellar activity cannot mimic the Ly$\alpha$ transit of 55 Cancari b, strengthening the conclusion that this planet has a partially transiting exopshere. We were able to compare our simulations to more active stars by artificially increasing the variability in the Solar Ly$\alpha$ light curve.  In the higher variability data, the largest value of \rp we measured is $<3\times$ the input value which again is not large enough to reproduce the Ly$\alpha$ transit depth reported for the more active stars HD 189733 and GJ 436, supporting the interpretation that these planets have extended atmospheres and possible cometary tails.
\end{abstract} 


\section{Introduction}\label{sec:intro}
Transiting exoplanets provide the opportunity to explore the atmospheric size and composition of planets outside our own Solar system. Observations in the near- and far-ultraviolet (NUV, FUV) are sensitive to the upper atmosphere of the planet where this radiation is absorbed. Lyman-$\alpha$ ($\lambda_{Ly\alpha}=121.6$ nm) emission of neutral hydrogen (H \textsc{i}) is the strongest chromospheric emission feature in the stellar FUV spectrum and the dominant source of UV flux for M dwarfs \citep{Ehrenreich:2011jra}.  

Indeed, direct observations of active M dwarfs have revealed a high FUV flux \citep{2014AJ....148...64S} and that the total Ly$\alpha$ flux can be up-to 75\% of the total UV flux (115 - 310 nm), far greater than the Solar value of 0.04\% \citep{France:2013gb}. Originating in the transition region between the chromosphere and corona, Ly$\alpha$ is used as a proxy for determining both the temperature and pressure profiles of the stellar atmosphere; however, observing Ly$\alpha$ emission from stars is challenging since it is entirely absorbed by the Earth's atmosphere and is heavily effected by absorption in the ISM requiring space-based observations of the nearest stars \citep{Linsky:1980dw}. 

Since Ly$\alpha$ emission is absorbed in the upper atmospheres of planets it offers one of the most promising opportunities to infer the presence of extended atmospheres of exoplanets. Deep Ly$\alpha$ transits as observed with HST/STIS have suggested the presence of extended atmospheres on the hot Jupiters HD 209458b, which orbits a Sun-like star \citep{VidalMadjar:2003bl,VidalMadjar:2004ir,Linsky:2010fo,2010ApJ...709.1284B} and HD 189733b which orbits a more active K1 dwarf \citep{LecavelierDesEtangs:2010kw,LecavelierDesEtangs:2012jq,Bourrier:2013gl}. Along with neutral hydrogen the presence of heavier elements have also been inferred in the upper atmosphere of HD 209458 in the FUV lines of O \textsc{i} (130.4 nm), C \textsc{ii} (133.5 nm), Si \textsc{iii} (120.6 nm), Si \textsc{iv} (139.8 nm), and Mg \textsc{i} (285.2 nm) \citep{VidalMadjar:2004ir,Linsky:2010fo,BenJaffel:2013ei,2010ApJ...722L..75S}. \citet{Poppenhaeger:2013wx} reported a tentative detection of HD 189733b in X-rays (0.1 - 6.3 nm) through multiple observations using \textit{Chandra}. Due to the difficulty of observing X-ray transits and the activity of the star \citet{Poppenhaeger:2013wx} find an X-ray transit depth between $2.4-9\%$, i.e. $1-3\times$ the optical transit depth. Further observations will help determine if the X-ray transit depth of HD 189733b is indeed deeper than in the optical, or if it is the result of stellar activity.
 
\citet{Ehrenreich:2012bq} reported a tentative detection of an extended atmosphere in Ly$\alpha$ of the warm Jupiter 55 Cancari b with a transit depth of $7.5\pm1.8\%$ detected with HST/STIS. This planet's transit has not been detected in optical  photometry suggesting that if additional Ly$\alpha$ absorption is caused by the planet then the extended atmosphere is grazing the stellar disk. Ly$\alpha$ observations of GJ 436b's transit using HST/STIS have also revealed atmospheric escape from the atmosphere of this hot Neptune orbiting an M dwarf. \citet{Kulow:2014bv} detected an absorption  signature at the end of the optical transit of GJ 436b with a transit depth of $\sim 25\%$, (far greater than the $0.69\%$ optical transit depth). A further study by \citet{Ehrenreich:2015ci} determined the ephemeris time used by \citet{Kulow:2014bv} to be inaccurate, and their reanalysis shifted this increased absorption signature into mid-transit. In their observations, \citet{Ehrenreich:2015ci} measured an even deeper asymmetric Ly$\alpha$ transit, with a depth of $56.3\pm3.5\%$ and found a cometary tail trailing the planet.

Heavy UV radiation from the host stars is thought to have ionized the upper atmospheres of these planets, resulting in hydrodynamical blow-off \citep{MurrayClay:2009ip,Tripathi:2015vh}, implying that these planets may have lost some of their atmosphere \citep{Batalha:2011fs,Pepe:2013hs,Hu:2015hc}. \citet{Rugheimer:2015da} studied the effects of Ly$\alpha$ on Earth-like planets orbiting M dwarfs and found that even when the Ly$\alpha$ flux is increased to 180$\times$ the largest observed value in the MUSCLES stellar sample \citep{France:2013gb} it only has a small effect on the photochemistry of the planet's atmosphere. However, \citet{Miguel:2015cz} showed that the impact of higher Ly$\alpha$ flux may drastically change the observable spectral features for mini-Neptunes in systems at lower pressures than those considered by \citet{Rugheimer:2015da}.
 
NUV observations of the heavily-irradiated hot Jupiter WASP-12b from HST/COS have revealed transit depths in the Mg {\sc ii} lines up-to $3\times$ deeper than the optical transit depth \citep{Fossati:2010do,Haswell:2012ft}. These transits also show an asymmetry in the timing of the transit, with their NUV (253.9-258.0 nm) events exhibiting an early ingress but ending simultaneously with the optical light curve. These deep transits have been attributed to the presence of heavy elements in the atmosphere of the planet; however, the early-ingress has also been attributed to the presence of a magnetospheric bowshock transiting ahead of the planet \citep{Vidotto:2010jha,Llama:2011de}.

Observations of HD 189733b showed no detectable Ly$\alpha$ transit in 2010 April but detected a $14.4\pm3.6\%$ transit depth in 2011 September \citep{LecavelierDesEtangs:2012jq}. This transit depth is $\sim 10\times$ deeper than the optical transit and has been attributed atmospheric blow-off from the planet. The changes in the transit depth are likely correlated to the variable stellar irradiation to which the planet is exposed. 
 
Given that the variability and large depths observed in Ly$\alpha$ transits are being used to infer properties of exoplanet atmospheres, it is necessary to ask whether such variability could be directly attributed to stellar activity affecting the transit observation and not the presence of an extended atmosphere on the planet. Indeed, stellar activity effects must be accounted for when analyzing transit observations, particularly at short wavelengths since the presence of stellar active regions can alter the inferred transit depth and potentially lead to false conclusions about the composition of the planetary atmosphere. In the optical, dark star spots on the stellar surface that intersect the transit chord will result in a shallower transit as the planet occults less star light as it passes over the star spot. The presence of star spots that do not intersect the transit chord results in a deeper transit (e.g., \citealt{Pont:2013ch}). 

\begin{figure} 
   \centering
   \includegraphics[width=0.5\textwidth]{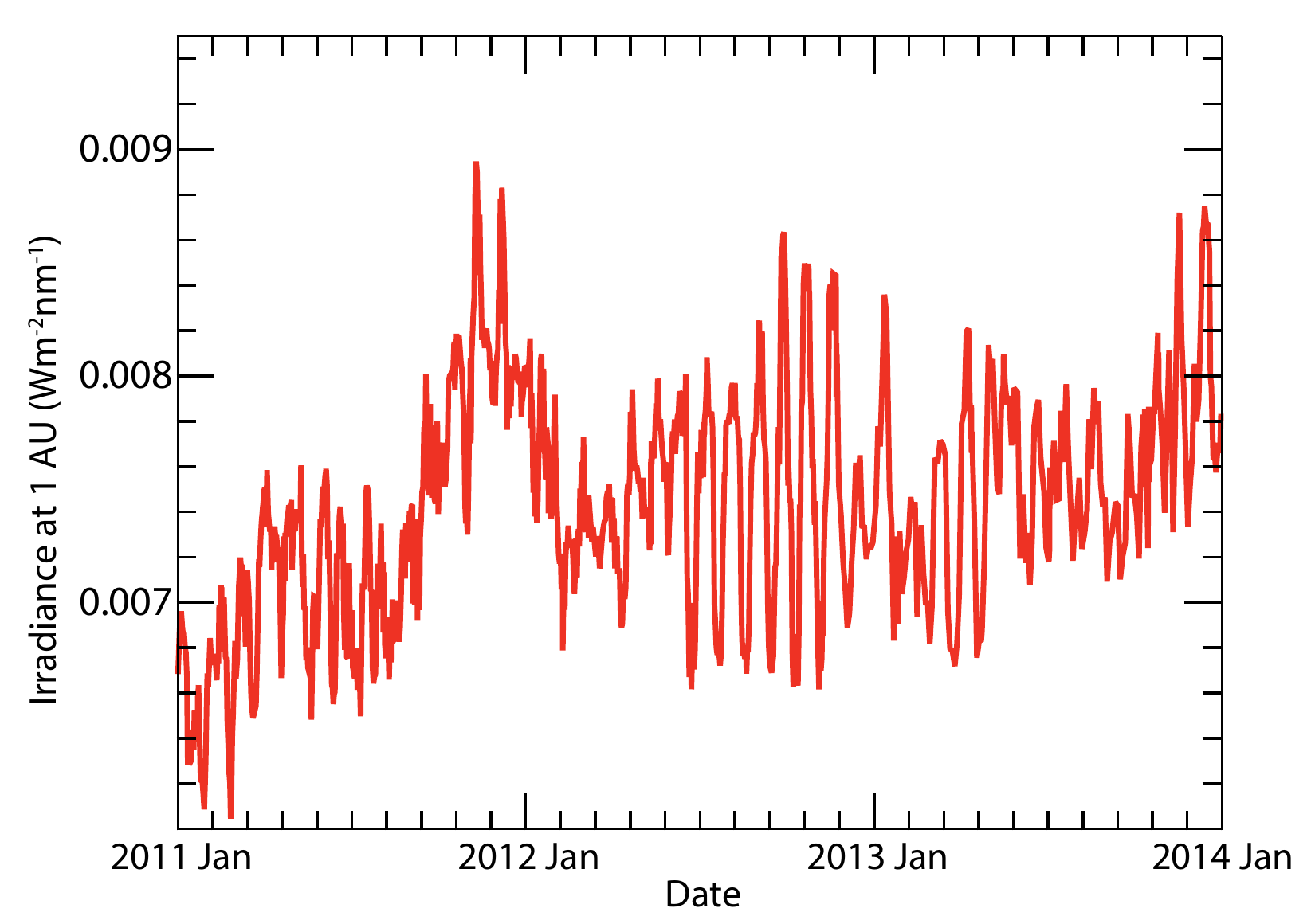} 
   \caption{Ly$\alpha$ irradiance (adjusted to 1 AU) of the Sun (smoothed to 1 h cadence) from 2011 January until 2014 January from the Extreme Ultraviolet Variability Experiment (EVE) instrument on board NASA's Solar Dynamics Observatory (SDO). The 27 day Solar rotation period can clearly be seen with variability of $\sim 30\%$, as well as the transition from Solar minimum to Solar maximum.}
   \label{fig:lyalc}
\end{figure} 
 
In \citet{Llama:2015bt}, we investigated how both occulted and unocculted active regions impact exoplanetary transits at soft X-ray, UV, and optical wavelengths. We simulated transits over disk-resolved images of the Sun with the aim of determining whether stellar activity can mimic an extended planetary atmosphere. At short wavelengths, the active regions appear as extended, bright patches on the stellar disk and therefore have the opposite effect than in the optical i.e., the occultation of a bright region along the transit chord results in a deeper transit and the presence of unocculted X-ray/UV bright regions on the stellar disk will result in a shallower transit. \citet{Llama:2015bt} showed that the impact of stellar activity was less prevalent at longer wavelengths (FUV and optical) due to the active regions being less extended than in the soft X-ray / EUV regime. 

\begin{figure*} 
\centering    
\includegraphics[width=1\textwidth]{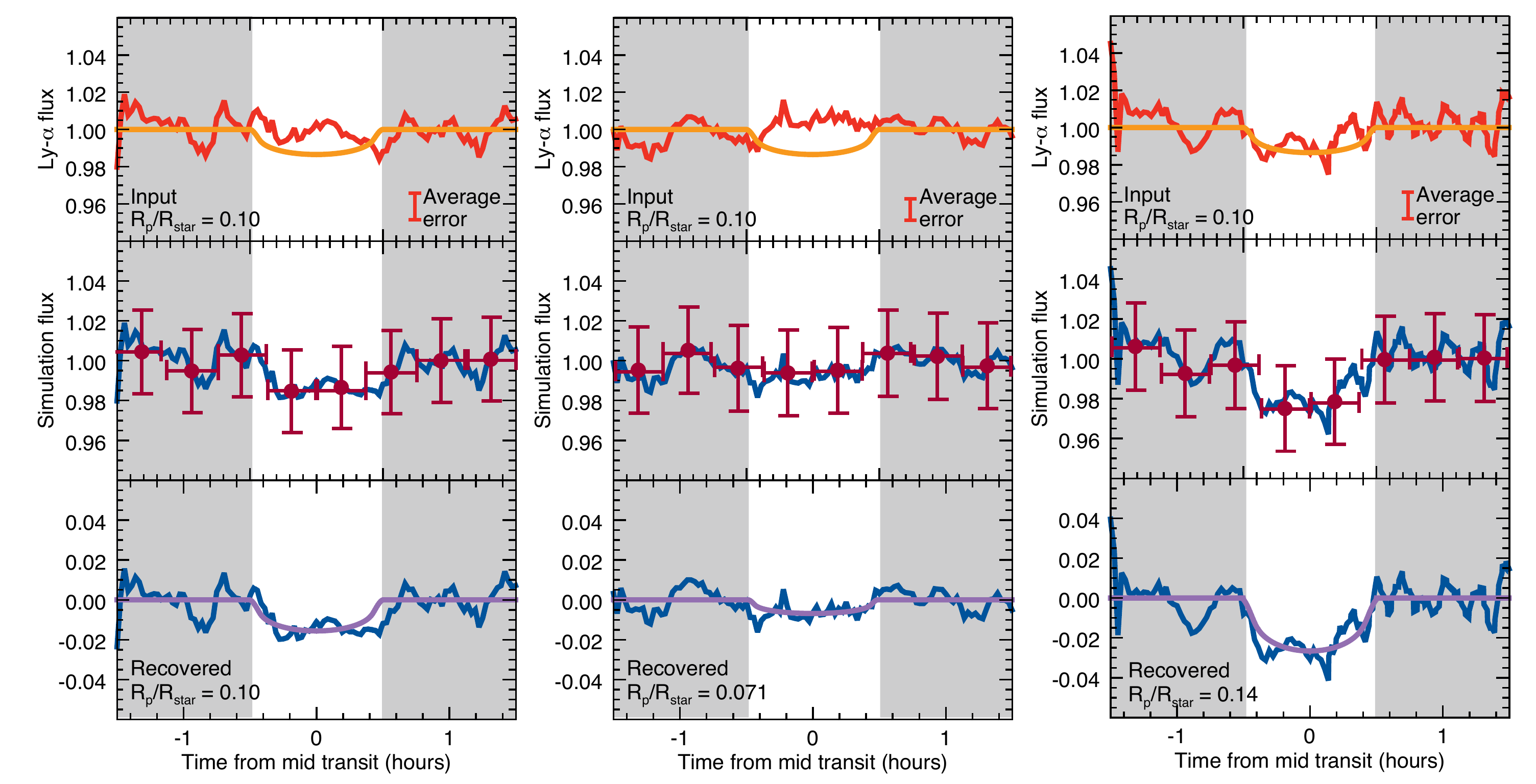} 
   \caption{Three of the 1100 simulations. The top panel shows the normalized 3 h Ly
  $\alpha$ light curve (red) and the input model transit light with \rp$=0.1$ curve (orange). The middle panel shows the combined Ly
  $\alpha$ and model transit light curve (blue). Over plotted is this light curve rebinned to a cadence similar to that of existing Ly
  $\alpha$ observations (see Section \ref{sec:lowresobs}). We then fit a transit model to find the best fit value of \rp (purple). The first column shows an example where we recover the input value of \rp, the middle column shows an example where we recover a smaller value of \rp, and finally an example where \rp is larger than the input value.}
\label{fig:lyasimulations}  
\end{figure*}

In this work, we used Ly$\alpha$ observations of the Sun taken continuously over 3.5 years to investigate the effect its variability has on our ability to recover the radius of a transiting exoplanet and aid the interpretations of Ly$\alpha$ transits of known exoplanets. We extracted 1100 continuous 3 h windows of high cadence (10 s) Ly$\alpha$ data into which we inserted exoplanet transits. This enabled us to determine the effect of stellar activity on a single planet transit observation in Ly$\alpha$, and to predict the number of transits needed to reliably measure the effective radius of the planet at this wavelength. To investigate the effects of stellar activity on more active stars, we scaled the Solar Ly$\alpha$ data to simulate light curves of higher variability stars such as HD 189733 and M dwarfs. 

\section{Solar observations}\label{sec:datamodel}

\subsection{SDO/EVE Observations}
NASA's \textit{Solar Dynamics Observatory} (SDO)\footnote{http://sdo.gsfc.nasa.gov} was launched in early 2010 and has been continuously monitoring the Sun at a range of wavelengths spanning the X-ray, UV, and optical. The Extreme Ultraviolet Variability Experiment (EVE)\footnote{http://lasp.colorado.edu/home/eve/} instrument on board SDO measures the disk integrated (Sun-as-a-star) EUV spectrum once every 10 s within the wavelength range 6.5-37 nm using the Multiple EUV Grating Spectrograph (MEGS-A) component and also between 37-105 nm with the MEGS-B component \citep{Woods:2010fw}. Additionally, the photo diode MEGS-P, covers the Ly$\alpha$ emission feature. Although the bandpass of MEGS-P is 10 nm wide, 99\% of the flux measured by this component is from the Solar Ly$\alpha$ emission \citep{Woods:2010fw}. To minimize the instrument degradation, MEGS-P only acquires data at a 10 s cadence for 3 h a day and for 5 min every hour the remainder of the time. 
 
Using the \textsc{idl} EVE routines from the Virtual Solar Observatory provided by SolarSoft\footnote{http://www.lmasal.com/Solarsoft}, we downloaded all the available EVE level 2 line data. These files contain the Solar irradiance (adjusted to 1 AU) at selected emission lines, including Ly$\alpha$. Fig.\,\ref{fig:lyalc} shows the Solar Ly$\alpha$ irradiance (as measured using MEGS-P) from 2011 January to 2014 January smoothed to a cadence of 1 hour. During this time the Sun was transitioning from Solar minimum toward maximum. 

There are two forms of variability present in this light curve. Firstly, long term rotational modulation caused by the 27 day Solar rotation period can clearly be seen. This modulation, whilst varying by up-to $\sim30\%$ is not problematic for transit observations, which are taken over the period of a few hours and so such long term variability will have an almost uniform effect on the light curve. Secondly, there is the presence of short term variability, caused by the evolution of the active regions themselves. This short term variability is more problematic when dealing with short observing windows since it affects the determination of the normalization level. It is the effect of this variability that degrades our ability to recover accurate values of the planet-to-star radius ratio, \rp. 
 
Also on board SDO is the Atmospheric Imaging Assembly (AIA). This instrument monitors the Sun in ten wavelengths ranging from soft X-ray/EUV through to the near ultraviolet and optical. Unlike EVE, this instrument records resolved images of the Solar disk with a spatial resolution of 1\arcsec\, on a CCD of 4096 px$^2$. These data were used in \citet{Llama:2015bt} to simulate exoplanet transit light curves using disk resolved data. We found that the presence of large, bright, unocculted active regions on the Solar surface affect the normalization of the simulated transit chord and makes accurately measuring \rp challenging in the soft X-ray/EUV regime. Here, we used disk integrated Ly$\alpha$ data and although we lose all spatial information, we provide the limits of probing the upper atmospheres of planets at key spectral signatures.

\begin{figure*} 
\centering        
\includegraphics[width=1\textwidth]{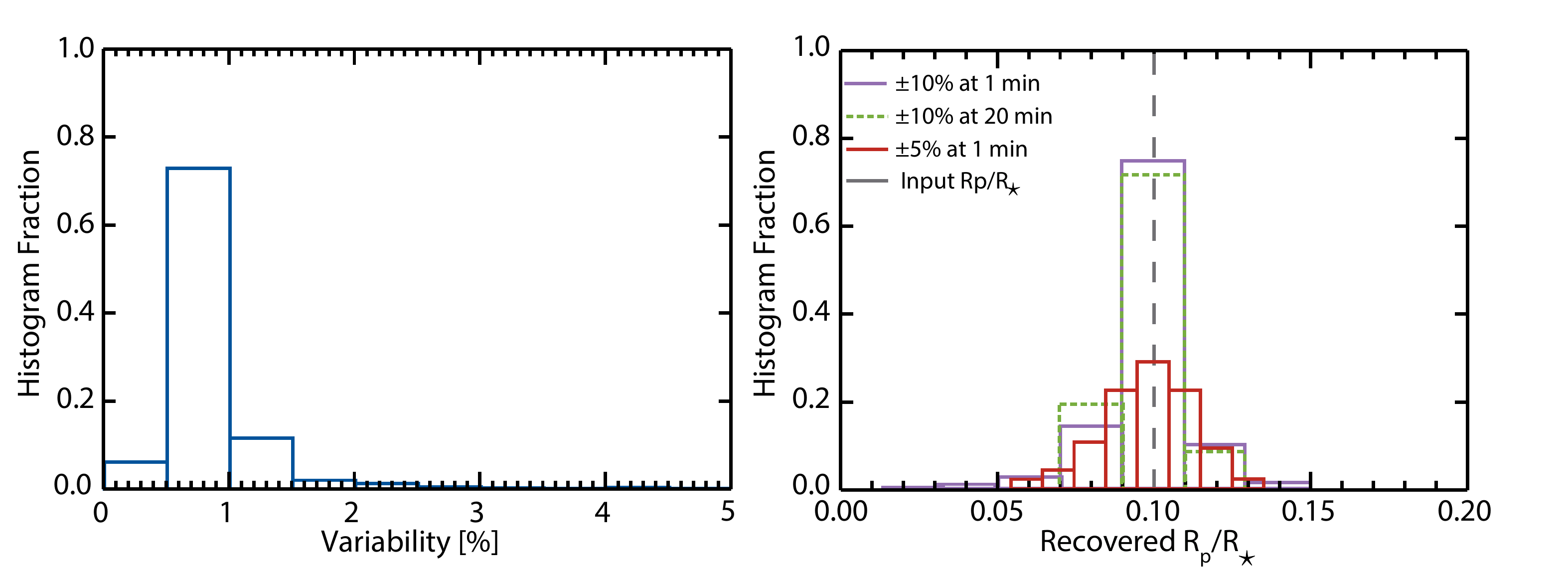} 
   \caption{\textit{Left:} Histogram showing the short-term variability, defined as $\sigma/\mu$ (per light curve). \textit{Right:} Histogram of measured \rp\, from the 1100 simulated exoplanet transits of a hot Jupiter with \rp$=0.1$ into the Solar Ly
  $\alpha$ light curves. The $\pm5\%$ bin width of the red histogram matches the binning presented in \citet{Llama:2015bt}, whilst the $\pm10\%$ bin width of the purple histogram matches typical Ly
  $\alpha$ and X-ray transits. The dashed green line shows the distribution of \rp for the lower cadence observations (see Section \ref{sec:lowresobs}).}
\label{fig:lyahist}  
\end{figure*}  
\section{Transit Model}\label{sec:tmodel}
The EVE MEGS-P Ly$\alpha$ data contains 3 h of continuous observations per day (at 10 s cadence), setting the duration limit of our simulated transit events. We began by isolating the 3 h of continuous data and rebinning it to 1 min cadence for each day in our data set. In total we were able to isolate 1100 separate 3 h windows of continuous Ly$\alpha$ data. Sample Ly$\alpha$ light curves are shown in the top panels of Fig.\,\ref{fig:lyasimulations} in red.

At the Ly$\alpha$ wavelength, the Sun is limb-darkened (e.g., \citealt{Curdt:2008dj}) and so we simulated the transit light curve of a hot Jupiter with \rp$=0.1$, at an impact parameter $b=0$ using the code developed by \citet{Mandel:2002bb}.  The 3 h simulated transit light is comprised of the 1 h transit event and 1 h either side of the transit. 

The simulated transit light curve is plotted in the top panels of Fig.\,\ref{fig:lyasimulations} in orange. The Ly$\alpha$ light curve is then multiplied together with the model transit light curve to produce the simulated data so that only the in-transit part of the Ly$\alpha$ light curve is affected. The solid line in the middle panels of Fig.\,\ref{fig:lyasimulations} shows the results of multiplying the light curves together. 

Using the simulated Ly$\alpha$ transit light curves we measured \rp by performing a $\chi^2$ test with simulated \citet{Mandel:2002bb} light curves with varying values of \rp to find the best fit transit light curve. For each case shown in Fig.\,\ref{fig:lyasimulations} the best fit transit is shown in purple in the bottom panels. The impact parameter in our fitted light curves is fixed to the input value of $b=0$ since this would be known from optical photometry. We also assumed the same limb-darkening coefficients as the input transit model, meaning the only free parameter in our fit is \rp.

\section{Results: Ly$\alpha$ Transit Variability}\label{sec:lyaresults}
We repeated the process outlined in Section \ref{sec:tmodel} and simulated 1100 transits of \rp$=0.1$ using each of the isolated 3 h windows of the Solar Ly$\alpha$ light curve. For each simulation we then recorded the measured value of \rp to quantify the effects of Ly$\alpha$ variability on our ability to measure \rp. The simulation shown in Fig.\,\ref{fig:lyasimulations} (left column) is an example of where the input value of \rp$=0.1$ was recovered successfully. In this case, the Ly$\alpha$ variability during the 3 h observing window was relatively small. The simulation in the middle column shows an example where the measured value of \rp is smaller than the input value, due to a large amount of short term variability in the light curve. The simulation in the right panel of Fig.\,\ref{fig:lyasimulations} is another example where there is significant short term variability in the Ly$\alpha$ light curve during the transit event, this time resulting in the best-fit model transit being one with a larger value of \rp. 
 
The left panel of Fig.\,\ref{fig:lyahist} shows a histogram of the variability present in the 3 h windows of Ly$\alpha$ Solar data. We use the ``coefficient of variance'', $\sigma/\mu$, where $\sigma$ is the standard deviation and $\mu$ is the mean of the 3 h data to compute the variability in the light curve. We find that during a 3 h window, the Ly$\alpha$ variability does not exceed 3\%, with 95\% of the windows showing less than 1.5\% variability. 

The right panel of Fig.\,\ref{fig:lyahist} shows a histogram of the measured values of the planet-to-star radius ratio, \rp, from the 1100 simulations. In all cases the input value was \rp$=0.1$. The red histogram has a bin width of $\pm5\%$ to match the results presented in \citet{Llama:2015bt}, whilst the bin size of the purple histogram was chosen to match the typical $\pm10$\% uncertainties of Ly$\alpha$ and X-ray transits. Both histograms peak at the correct value of \rp$=0.1$ (within error). For the larger uncertainty histogram (purple), we find the correct answer is found with a frequency of $\sim75\%$, whilst for the smaller uncertainty histogram (red), we find the input value with a frequency of $\sim30\%$. The maximum value of \rp in our simulations is 50\% larger than the input value which cannot explain the large transit depths for both HD 189733b and GJ 436b \citep{Bourrier:2013gl,Kulow:2014bv,Ehrenreich:2012bq}. 

\section{discussion}
\subsection{55 Cancari b: The transit of a planetary exosphere}
\citet{Ehrenreich:2012bq} reported a tentative detection of 55 Cancari b, a planet previously only found in radial velocity. They reported a Ly$\alpha$ transit depth of $7.5\pm1.8\%$, suggesting that if this absorption is from the planet, then this system may have a partially transiting exosphere. Using our Ly$\alpha$ light curves we were able to test this theory. We repeated the simulations as described in Section \ref{sec:tmodel}, this time without inserting an exoplanet transit. We then measured the best-fit \rp value to determine whether stellar activity alone can mimic a transit event. From the 1100 simulations, 65\% of the light curves are best fit with no transit, and  the median value of \rp $=0.02$, much smaller than the Ly$\alpha$ depth reported for 55 Cancari b by \citet{Ehrenreich:2012bq}. This finding strengthens the conclusion that the increased Ly$\alpha$ absorption that coincides with the transit of 55 Cancari b is unlikely caused by stellar activity but by the planet having an extended atmosphere that is partially transiting the stellar disk.  

\subsection{Comparison with disk resolved data}
\begin{figure} 
\centering    
\includegraphics[width=0.5\textwidth]{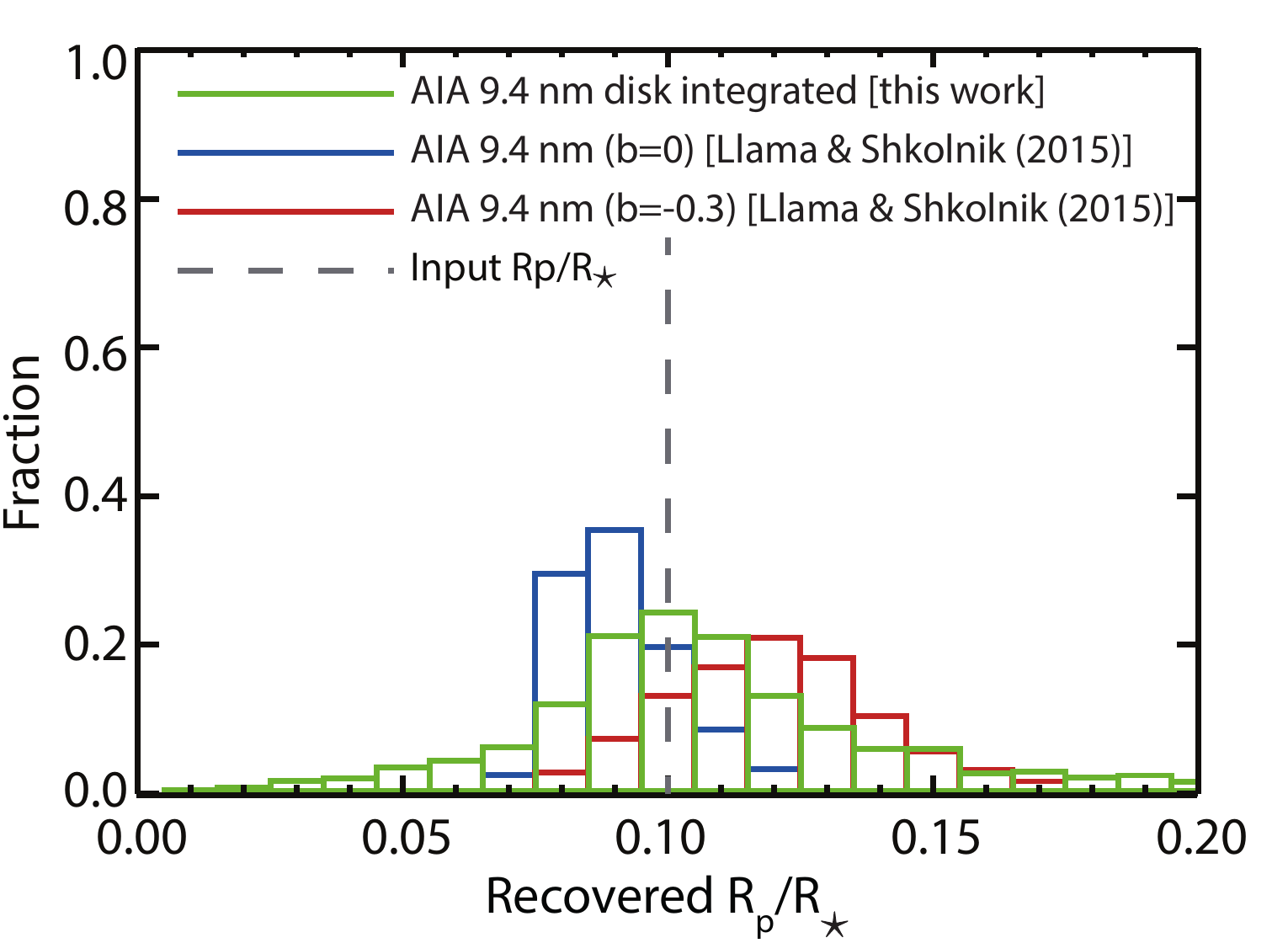} 
   \caption{Histogram showing the distribution of measured \rp using the disk integrated AIA 9.4 nm data (green). Over plotted is the measured \rp from the transit simulations over the disk resolved AIA 9.4 nm data as shown in Fig. 7 of \citet{Llama:2015bt}. }
\label{fig:aiacomp}  
\end{figure}

In \citet{Llama:2015bt} we were able to make use of disk resolved images of the Sun obtained by AIA and account for both occulted and unocculted active regions in the simulated light curves, rather than the disk-integrated data as collected by EVE. We carried out two sets of simulations  over the ten wavelengths observed by AIA, one where the planet transited over the activity belts of the Sun ($b=-0.3$), and one where the planet transited over the equator ($b=0$). In the first case, on average we measured a larger value of \rp than the input value due to the planet frequently occulting active regions. In the latter case, the presence of unocculted active regions on the stellar surface meant the transit depth was on average shallower, resulting in a smaller value of \rp being measured. The blue and red histograms shown in Fig.\,\ref{fig:aiacomp} show the measured \rp distribution for the soft X-ray (9.4 nm) $b=0$ and $b=-0.3$ simulations respectively. 

Multiple studies of the  HD 189733 system have confirmed that the lack of bumps (caused by the planet occulting a star spot) in the transit light curve indicates it is unlikely that the transit chord is intersecting the activity belts of the star \citep{Pont:2013ch,Poppenhaeger:2013wx}. 
 
Here, we do not have information on the locations of the star spots; however, the EVE data set also contains the same EUV data from AIA but as disk-integrated rather than disk-resolved data. This allows us, at least for the EUV wavelengths, to directly compare the simulation method used here with that presented in \citet{Llama:2015bt}. It is worthy of note that the 9.4 nm soft X-ray images of the Sun are limb-brightened rather than limb-darkened, meaning the transit light curve exhibits a W-shape rather than a U-shape. We therefore used a limb-brightened transit model (originally developed by \citet{2010ApJ...722L..75S}) rather the limb-darkened model shown here. For a full description of the limb-brightened model please see Section 3.3 of \citet{Llama:2015bt}.

The green histogram in Fig.\,\ref{fig:aiacomp} shows the distribution of measured \rp using the disk integrated AIA 9.4 nm data and the simulation method presented in Section \ref{sec:tmodel} (but using a limb-brightened transit model). The histogram shows that the spread in \rp is more symmetric about the correct value of \rp$=0.1$ for the disk integrated data and has a similar spread to the results where we used disk resolved data; however, since we were unable to simulate the effects of the planet directly occulting active regions this method provides a lower limit to the effects of activity on Ly$\alpha$ transit observations.

\begin{figure} 
\centering     
\includegraphics[width=0.5\textwidth]{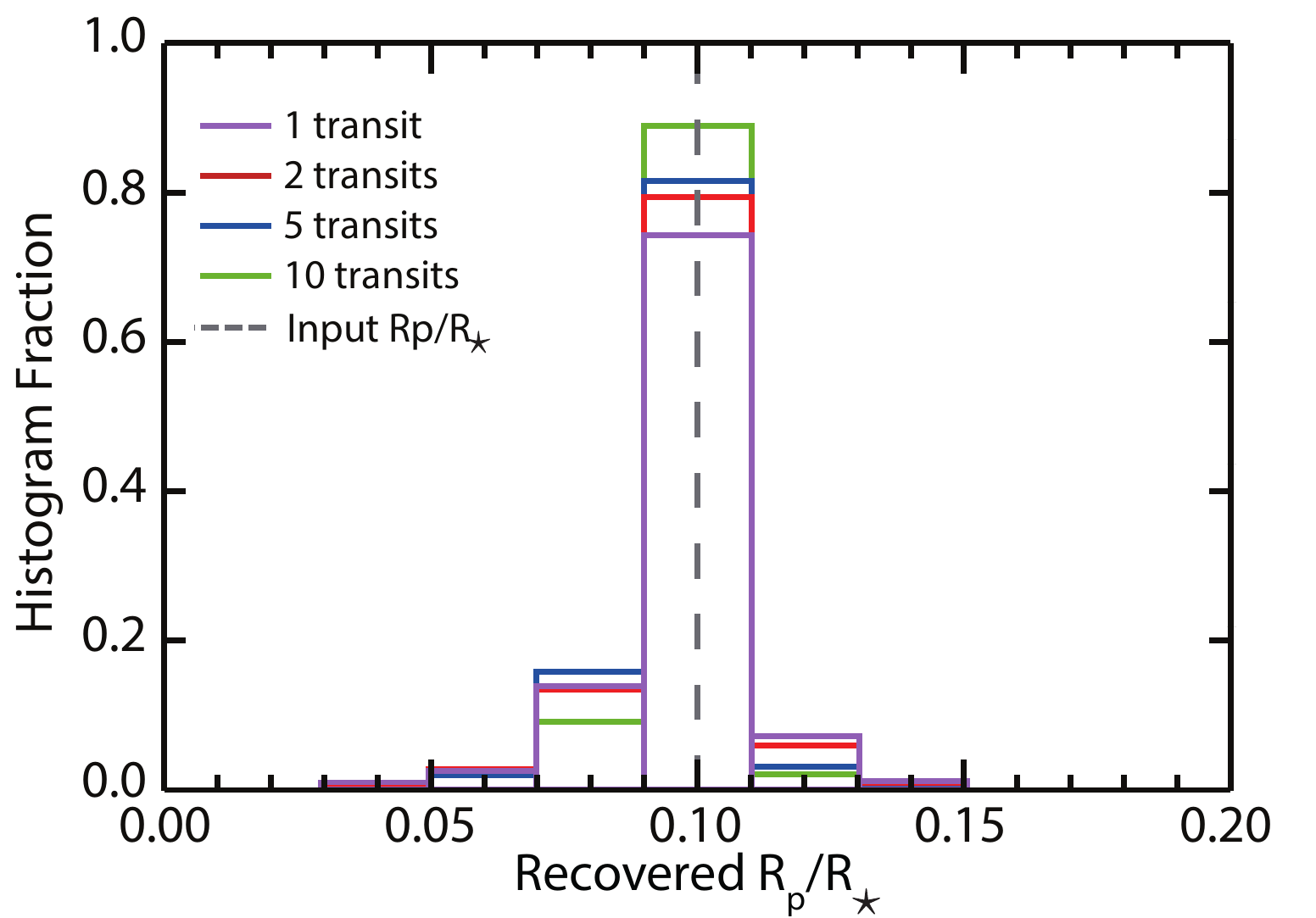} 
   \caption{Histogram showing the distribution of measured \rp from combining 2, 5, and 10 separate transits together. Combining transits together improves the distribution of \rp with the correct answer being obtained in 90\% of the simulations once more than 10 transits are combined.}
   \vspace{0.1in}
\label{fig:ntransits}  
\end{figure} 

\subsection{Lower cadence of observation}\label{sec:lowresobs}
The histogram of measured \rp shown in Fig.\,\ref{fig:lyahist} was created using the Ly$\alpha$ light curve binned to a 1 min cadence. Observations of exoplanet transits require longer integration times to obtain sufficient signal-to-noise and so are comprised of fewer data points. \citet{Llama:2015bt} found that for the lower cadence light curves, the soft X-ray and EUV transit depths were on average shallower than in the high cadence case. To investigate the impact of longer integration times on the simulations presented here, we rebinned our 1100 Ly$\alpha$ transit light curves to a cadence of 20 min so that each transit event (including out-of-transit) was comprised of 9 data points, rather than 180 at 1 min cadence. Sample rebinned light curves are shown as data points in the middle panel of Fig.\,\ref{fig:lyasimulations}. For each of these rebinned light curves we also measured the best fit value of \rp. We found a similar trend to that reported by \citet{Llama:2015bt}, where the measured value of \rp is lower than the full cadence simulations. The dashed histogram in Fig.\,\ref{fig:lyahist} shows the distribution of measured \rp for the lower cadence simulations. 
 
\begin{figure*}   
\centering     
\includegraphics[width=1\textwidth]{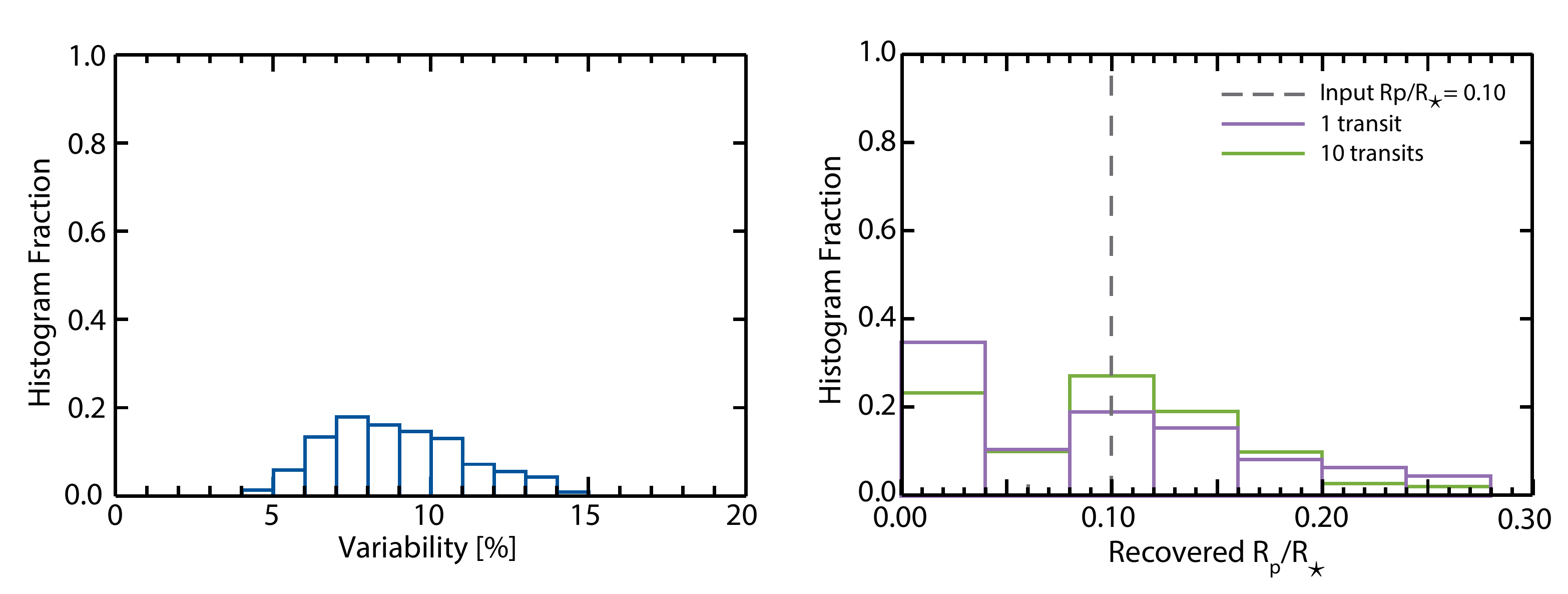}  
   \caption{\textit{Left:} Histogram showing the variability, defined as $\sigma/\mu$ (per light curve), for the simulated higher activity Ly$\alpha$ data. Now the variability ranges from $5-15\%$, similar to the variability levels reported for HD 189733 and GJ 436 \citep{BenJaffel:2013ei,Ehrenreich:2015ci}. \textit{Right:} Histogram of measured \rp for the 1100 light curves using the simulated higher activity data (purple). For the higher amplitude data, we find $\sim35\%$ of the simulated light curves are best fitted with no transit. By combining 10 or more transits (green histogram) we find that the histogram then peaks at the correct answer of \rp$=0.1$. }
\label{fig:lyaf6}   
\end{figure*}
\subsection{Combining data from multiple transits}
\citet{Poppenhaeger:2013wx} observed X-ray ($0.1-6.3$ nm) transits of HD 189733b using \textit{Chandra}. By combining multiple transits they were able to determine an X-ray transit depth between $2.4-9\%$, i.e, $1-3\times$ deeper than the optical transit. Between these observations the distribution of star spots on the surface of the star will have changed, which will impact the transit recovery. To test how stellar activity will impact multiple transit observations we carried combined multiple Ly$\alpha$ light curves and measured the resultant \rp value. Fig.\, \ref{fig:ntransits} shows the results of combining multiple transits, which are chosen at random and then measuring the value of \rp. For 2 transits, the correct value of \rp is measured in $80\%$ of the simulations, and after 10 transits the correct value is obtained in $90\%$ of the cases. We therefore conclude that despite stellar activity changes on the surface of the star, combining multiple transits increases the likelihood of measuring the correct value of \rp, at least 90\% of the time.

\subsection{Simulating more active stars}
\citet{Bourrier:2013gl} reported a $14.4\pm3.6\%$ depth in the Ly$\alpha$ transit of HD 189733b in 2011 April, which is $\sim10\times$ deeper than the optical transit. \citet{BenJaffel:2013ei} reported an early ingress and increased absorption at multiple FUV wavelengths during transits of HD 189733b using HST/COS. They analyzed the O \textsc{i}, C \textsc{ii}, Si \textrm{iv}, Si \textrm{iii} lines. In their O \textsc{i} light curve they found a transit depth consistent with \rp =1.7$\times$ the optical radius. \citet{BenJaffel:2013ei} carried out a short-term analysis of the activity on HD 189733 using the O \textsc{i} line in their data. They found that during their 5 h observing campaign the FUV variability of HD 189733 was $\sim5\%$. \citet{Ehrenreich:2015ci} carried out an analysis of the unabsorbed red-wing of the Ly$\alpha$ line in their GJ 436 observations and estimated the activity to vary by 5-11\% during a single visit. 
 
From our histogram of short term Ly$\alpha$ variability (left panel of Fig.\,\ref{fig:lyasimulations}) we find the Sun is less variable than HD 189733, with 95\% of our Ly$\alpha$ light curves showing less than 1.5\% variability. To investigate the effect of a higher level of variability on our simulations we scaled the Solar Ly$\alpha$ light curve to attain variability levels comporabale to HD 18733 and GJ 436. To achieve this we took the normalized Ly$\alpha$ flux for each of the 1100 3 h sets of observations and raised it to the tenth power. We note that this does not change the time scale of the variability, it only increases the overall amplitude of the variability. The left panel of Fig.\,\ref{fig:lyaf6} shows the new variability distribution for the higher variability data, again defined as $\sigma/\mu$. The variability now ranges between$\sim5-15\%$, similar to the short term variability estimates for HD 189733 and GJ 436.

The right panel of Fig.\,\ref{fig:lyaf6} shows the distribution of measured \rp for this data set obtained using the same method presented in Section \ref{sec:tmodel}. The purple histogram shows the distribution of \rp for a single observation has a larger spread in measured values of \rp than for the Solar variability data set. Almost 40\% of the simulations are best fitted the no transit model, i.e, the variability is much larger than the transit depth.

In all 1100 of the higher variability simulations, we did not measure a value of \rp large enough to produce the $56.3\%$ transit depth reported for GJ 436b, again suggesting that the large Ly$\alpha$ transit depth of GJ 436 detected by \citet{Kulow:2014bv} and \citet{Ehrenreich:2015ci} is unlikely to be a mimic of stellar activity but rather supports the conclusion that this planet hosts an extended atmosphere.  We also combined data sets to investigate the value of repeated transit observations. We find a large number of transits ($>10$) are required before the histogram peaks at the correct value of \rp$=0.1$.  We emphasize that the results presented in this Section are based on artificially boosting the variability in the Solar Ly$\alpha$ light curve and do not alter the time scale of the variability.
  
\section{Summary}
Ly$\alpha$ transit observations provide the opportunity to probe the outermost atmospheres of exoplanets. However, at these wavelengths the impact of stellar activity is more prevalent than in the optical and infra-red and so must be accounted for when analyzing X-ray and UV observations to ensure an accurate measurement of the planet-to-star radius is obtained. At these high-energy wavelengths, active regions appear much brighter than their surroundings and evolve on a faster timescale than star spots in the optical.

Here we have used the Solar Ly$\alpha$ light curve obtained through the EVE instrument on board NASA's SDO spacecraft and inserted synthetic transits in an effort to determine how reliably we can measure the input planetary radius in the presence of stellar activity. We found that for stars of similar activity levels to that of our Sun, $\sim75\%$ of Ly$\alpha$ transits will recover the true depth of the transit (to within $\pm10\%$). We investigated the effect of longer integration times and found that on average the measured value of \rp will be smaller due to active regions being confined areas on the stellar disk; however, this effect is relatively small. For stars more active than the Sun, such as HD 189733 and most M dwarfs, our results serve as lower limits to the extent to which stellar activity will impact the ability to accurately recover the true planet-to-star radius ratio in Ly$\alpha$ observations. 

We also ran simulations without inserting an exoplanet transit to test whether stellar activity alone can mimic an exoplanet transit. We found that 65\% of the simulated light curves were best fit with no transit. In the remaining light curves we found no cases where the transit depth was comparable to the large depth found by \citet{Ehrenreich:2012bq} that coincides with the transit of 55 Cancari b. This strengthens the conclusions that this planet does indeed have a partially transiting exosphere.

By artificially boosting the amplitude of the variability in the Solar Ly$\alpha$ light curve to the levels reported for HD 189733 and GJ 436 we found that $40\%$ of the simulations are best fitted with no transit and only after 10 observations does the distribution of \rp peak at the correct value.  However, we did not measure a value large enough to explain the large Ly$\alpha$ transit depths reported for GJ 436b by \citet{Kulow:2014bv} and \citet{Ehrenreich:2015ci}, supporting the conclusion that this planet does indeed have an extended atmosphere. The findings and analyses presented here will be of particular importance when attempting to characterize the atmospheres of potentially habitable planets around M dwarfs. 

\acknowledgements 
The authors wish to thank the anonymous referee for their insightful comments that helped to improved the manuscript. We also wish to thank Andrew Collier-Cameron (St Andrews) for useful discussion. This work is supported by NASA Origins of the Solar System grant No. NNX13AH79G. JL also acknowledges support from STFC grant ST/M001296/1. The Solar Dynamics Observatory was launched by NASA on 2010 February 11, as part of the Living With A Star program. The EVE instrument EVE was built at the University of Colorado, Boulder in the Laboratory for Atmospheric and Space Physics (LASP), the University of Southern California (USC) and the Massachusetts Institute of Technology, Lincoln Laboratory (MIT-LL).

\bibliography{eve_v1}  
\end{document}